\documentclass[useAMS,usenatbib,twocolumn]{mnras}
\usepackage{epsfig,amssymb}
\usepackage{graphicx,amsmath,multicol}
\usepackage{multirow}
\usepackage{caption}
\usepackage{subcaption}

\usepackage{placeins}

\usepackage{physics}
\usepackage[T1]{fontenc}
\usepackage{aecompl}

\usepackage[utf8]{inputenc}
\usepackage{setspace}
\usepackage{graphicx}
\usepackage{mathtools}
\usepackage{grffile}
\usepackage{amsmath}
\usepackage{longtable}
\usepackage{graphics}
\usepackage{verbatim}
\usepackage{float}

\def\apj{{ ApJ}}
\def\apjl{{ApJL}}
\def\apjs{{ ApJS}}
\def\apss{{ Ap\&SS}}
\def\aap{{ A\&A}}

\def\mnras{{ MNRAS}}

\def\aa{{ A\&A}}

\def\nat {{ Nature}}

\def\ssr{{ Space Sci. Rev.}}

\def\PRE{{ Phys. Rev. E}}
\def\prd{{ Phys. Rev. D}}
\def\PRL{{ Phys. Rev. Lett}}

\def\jcap{{JCAP}}
\def\mpla{{MPLA}}
\def\pasp{{PASP}}

\title[Magnetized white dwarf cooling]
{Luminosity and cooling of highly magnetized white dwarfs: suppression of luminosity by strong magnetic fields}

\author[Bhattacharya, Mukhopadhyay \& Mukerjee]
{Mukul Bhattacharya,$^{1}$ 
Banibrata Mukhopadhyay$^{2}$\thanks{E-mail: mukul.b@utexas.edu,bm@iisc.ac.in,smukerjee@iisc.ac.in}
and Subroto Mukerjee$^{2}$\\  
$^{1}$ Department of Physics, University of Texas at Austin, Austin, TX 78712, USA\\
$^{2}$ Department of Physics, Indian Institute of Science, Bangalore 560012, India}
\begin{document}

\date{Accepted . Received ; in original form }

\pagerange{\pageref{firstpage}--\pageref{lastpage}} \pubyear{2018}

\maketitle

\label{firstpage}

\begin{abstract}
We investigate the luminosity and cooling of highly magnetized white dwarfs with electron-degenerate cores and non-degenerate surface layers where cooling occurs by diffusion of photons.
We find the temperature and density profiles in the surface layers or envelope of white dwarfs
by solving the magnetostatic equilibrium and photon diffusion equations in a Newtonian framework. We also obtain the properties of white dwarfs at the core-envelope interface, when the core is
assumed to be practically isothermal. With the increase in
magnetic field, the interface temperature increases whereas the interface 
radius decreases. For a given age of the white dwarf and for fixed interface radius or interface temperature, we find that the luminosity decreases significantly from about $10^{-6}\, L_{\odot}$ to $10^{-9}\, L_{\odot}$ as the magnetic field strength increases from about $10^9$ to $10^{12}\,$G at the interface and hence the envelope. This is remarkable because it argues that magnetized white dwarfs are fainter and can be practically hidden in an observed Hertzsprung--Russell diagram. We also find the cooling rates
corresponding to these luminosities. Interestingly, the decrease in temperature with time,
for the fields under consideration, is not found to be appreciable.

\end{abstract}


\section{Introduction}
\label{sec1}

One of the most puzzling observations in high energy astrophysics in the last decade or so is that of the overluminous Type~Ia supernovae. More than a dozen such supernovae have been observed since 2006 (see e.g. \citealt{Howell, Scalzo}). Their significantly high luminosities can only be explained if we invoke very massive progenitors, of mass $M\ge 2M_\odot$.
Proposed models to explain these highly super-Chandrasekhar progenitors include rapidly (and differentially) rotating white dwarfs \citep{Yoon} and binary evolution of accreting differentially rotating white dwarfs \citep{Hachisu}. Another set of proposals that has recently brought the issue of super-Chandrasekhar white dwarfs into the limelight relates to highly magnetized
white dwarfs. In a series of papers, the main message of this work, initiated by our group, has been that the enormous efficiency of a magnetic field, irrespective of its nature of origin, quantum (owing to constant super-strong field, e.g.
\citealt{dm12,dm13,dmr13}),
classical and/or general relativistic (owing to a varying strong field exerting magnetic pressure and tension: 
e.g. \citealt{dm14,Sathya}), can explain the existence of significantly super-Chandrasekhar white dwarfs
 (see e.g. \citealt{mukhoall}, for the current state of this research). 
 
Remarkably, unlike other proposals, this work also adequately predicts the required mass range $2.1 < M/M_{\odot} < 2.8$ of the progenitors in order to explain the set of overluminous Type~Ia supernovae. 
Note interestingly that observations \citep{ferra} indeed confirm that highly magnetized white dwarfs ($B$ $\gtrsim 10^{6}\,$G) are more massive than non-magnetized white dwarfs. The impact of high magnetic fields not only lies in increasing the limiting mass of white dwarfs but it is also expected to change other properties including luminosity, temperature, cooling rate etc. 
For example, poloidally dominated magnetized white dwarfs are shown to be smaller in size (e.g. \citealt{dm15,Sathya}). This can account for their lower luminosity, provided their surface 
temperature is similar to or lower than their corresponding non-magnetic 
counterparts.

Although magnetized white dwarfs, with fields much weaker than those considered by our group, 
were explored earlier (e.g. \citealt{Ost-hart,Adam}), 
nobody concentrated on the effects of magnetic fields on the internal properties such as thermal conduction, cooling rate, luminosity, etc. 
However, these effects become important when the chosen field strength is 
comparable to or larger (see e.g. \citealt{Adam}) than the critical 
field $B_{\rm c}=4.414\times 10^{13}\,$G, at which the Compton wavelength of the
electron becomes comparable to the corresponding cyclotron wavelength.
Super-Chandrasekhar, magnetized white dwarfs were also explored with relatively weaker central fields around $5\times 10^{14}\,$G, where the underlying magnetic pressure gradient, determined by the field geometries and profiles, is responsible for making the mass super-Chandrasekhar \citep{dm14,dm15,Sathya}.
All these magnetized white dwarfs appear to have multiple implications
(e.g. \citealt{mukhrao,arsco}), 
apart from their possible link to peculiar over-luminous Type~Ia supernovae.
Hence, their other possible properties must be explored.

Here in an exploratory manner, 
we estimate the luminosities of magnetized white dwarfs and 
calculate the corresponding cooling. This has become more relevant as magnetized
white dwarfs have been proposed to be candidates for soft gamma-ray
repeaters and anomalous X-ray pulsars, with ultraviolet luminosities too
small to detect \citep{mukhrao}. Also the white dwarf pulsar AR Sco has been very recently argued to be a
proto--highly magnetized white dwarf \citep{arsco}. While the cooling of white dwarfs is not a completely resolved issue, it has been investigated since the 1950s, when \citet{Mestel} attempted to understand the source of energy of white dwarfs and to estimate the ages of observed white dwarfs. Subsequently,
the cooling of white dwarfs was 
explored by \citet{MestelRud} and white dwarfs were found to be radiating at the expense of their thermal energy. 
The evolution and cooling of low--mass white dwarfs, beginning as a bright central star to the stage of crystallization after about $10\,$Gyr, were also addressed \citep{Tutukov} and it was argued that the similarity of a modern cooling curve to the one predicted by \citet{Mestel} is the 
consequence of a series of accidents. Indeed, the limitations of  
Mestel's original theory, and underlying approximations for white dwarf cosmochronology, were mentioned later \citep{font},
without undermining the essential role played by the theory for the historical development 
of the field of white dwarfs.
Furthermore, the physics of cool white dwarfs was reviewed \citep{Hansen}, with particular attention to their usefulness to extract valuable information about the early history of our Galaxy.

The above work either did not consider the effects of magnetic field or the fields embedding the star were assumed to be too weak to have any practical effects. On the other hand, the field of magnetized white dwarfs considered by our group (and some others) is higher than that of all previous work that addressed the cooling of white dwarfs. Hence, here we explore the luminosity and cooling of magnetized white dwarfs.

This paper is organized as follows. 
In section \ref{sec2}, we include the contribution of the magnetic field to the pressure, density, opacity and equation of
state (EoS) of white dwarfs and compute the 
resultant density and temperature profiles in envelope for different luminosities and magnetic field strengths. 
Subsequently, in section \ref{sec3}, we consider white dwarfs having either a fixed interface radius or a fixed interface temperature and evaluate 
their luminosities for increasing field strengths. In section \ref{sec4}, we compute the cooling rates of magnetized white dwarfs 
for the cases discussed in section \ref{sec3}. Next, we discuss the implications of our results for magnetized white dwarfs in section \ref{sec5} and we conclude with a summary in section \ref{sec6}.

\section{Temperature profile for a magnetized white dwarf}
\label{sec2}
In this section, we solve the magnetostatic equilibrium and photon diffusion equations in the presence of a magnetic field ($\vec{B}$) 
and investigate the temperature profile inside a white dwarf. We mainly perform our calculations for radially varying magnetic fields that are realistic. 
The presence of $\vec{B}$ inside a white dwarf gives rise to a magnetic 
pressure, $P_{B} = {B^2}/{8\pi}$, where $B=\sqrt{\vec{B}.\vec{B}}$, which contributes 
to the matter pressure to give rise to the total pressure (see, e.g., \citealt{Sinha}). Furthermore, the density also has a contribution from the magnetic field that is given by $\rho_{B} = {B^2}/{8\pi c^2}$ \citep{Sinha}. $\vec{B}$ also modifies the opacity and EoS of the matter therein.
Such a situation can be tackled more ingeniously in the general relativistic framework rather than Newtonian framework. 
Nevertheless, here, as a first approximation, we construct the magnetostatic equilibrium and photon diffusion equations in a Newtonian 
framework as
\begin{equation}
\frac{d}{dr}(P+P_{B}) = -\frac{GM}{r^2}(\rho+\rho_{B}),
\label{eqn1}
\end{equation}
and
\begin{equation}
\frac{dT}{dr} = -\frac{3}{4ac}\frac{\kappa(\rho+\rho_{B})}{T^{3}}\frac{L}{4\pi r^2},
\label{eqn2}
\end{equation}
respectively, neglecting magnetic tension terms. In these equations, $P$ is 
the matter pressure which is same as the electron degeneracy pressure in the core,
$\rho$ is the density of matter, $\kappa$ is the radiative opacity, $T$ is the temperature, $a$ is the radiation constant, $c$ is the speed of light in vacuum, $G$ is Newton's gravitational constant, in the envelope 
$m(r) \approx M$ is the mass enclosed within radius $r$, and $L$ is the luminosity. 

\begin{figure*}
  \begin{subfigure}[tp]{0.5\linewidth}
    \centering
    \includegraphics[width=0.97\linewidth]{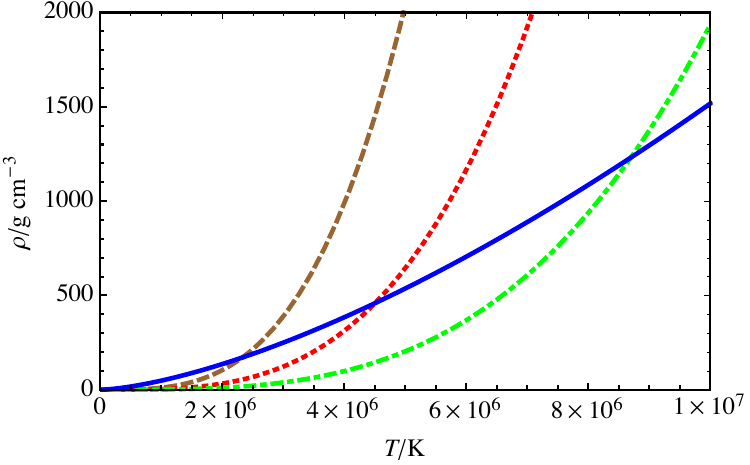}
  \end{subfigure}
  \begin{subfigure}[tp]{0.5\linewidth}
    \centering
    \includegraphics[width=0.97\linewidth]{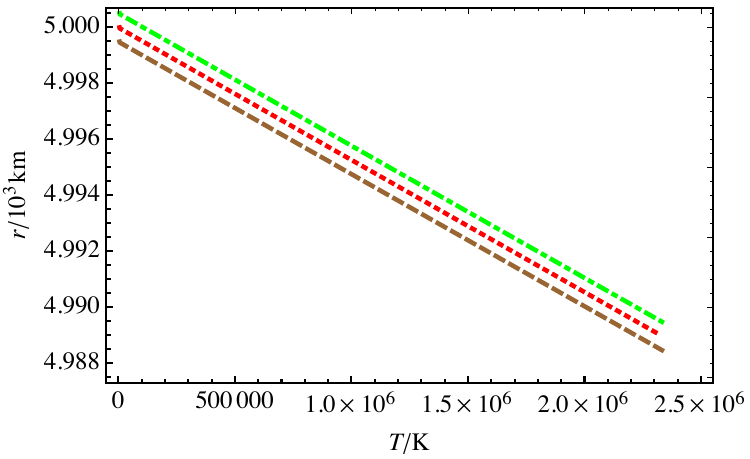}
  \end{subfigure}  
  \caption{
 {\it Left-hand panel:} variation of density with temperature for non-magnetized white dwarfs with $L$: $10^{-5} L_{\odot}$ (dashed line), 
$10^{-4} L_{\odot}$ (dotted line) and $10^{-3} L_{\odot}$ (dot-dashed line). The $\rho_{*}$ and $T_{*}$ are obtained from the intersection of the $\rho-T$ profiles with equation (\ref{eqn4}) (solid line).
 {\it Right-hand panel:} variation of radius with temperature for non-magnetized white dwarfs with $L$: $10^{-5} L_{\odot}$ (dashed line), 
$10^{-4} L_{\odot}$ (dotted line) and $10^{-3} L_{\odot}$ (dot-dashed line). The $r$-axis is rescaled by a factor of 0.9999 (1.0001) for $L = 10^{-5} L_{\odot}$ ($10^{-3} L_{\odot}$) to avoid overlap.
 }
  \label{fig1} 
\end{figure*}

The opacity for a non-magnetized white dwarf is approximated with Kramers' formula, $\kappa = \kappa_{0}\rho T^{-3.5}$, where $\kappa_{0} = 4.34\times10^{24} Z (1+X)\,$$\rm{ cm^{2}g^{-1}}$ and $X$ and $Z$ are the mass fractions of hydrogen and heavy elements (elements other than hydrogen and helium) in the stellar interior, respectively \citep{Schwarzschild}. For a typical white dwarf, $X=0$, and we assume for simplicity the mass fraction of helium $Y=0.9$ and $Z=0.1$. The opacity is due to the bound-free and free-free transitions of electrons \citep{Shapiro}. For the typically large $B$ considered in this work, the variation of radiative opacity with $B$ can be modelled similarly to neutron stars as $\kappa = \kappa_{B} \approx 5.5\times10^{31} \rho T^{-1.5}B^{-2}\,$$\rm{cm^{2}g^{-1}}$ \citep{PotYak,VenPot}. 
Note that across the surface layers of the white dwarf, radiation conduction dominates over the electron conduction and hence the same goes with the corresponding opacities \citep{PotYak}.

It has already been shown that if we include the effects of a magnetic pressure gradient and magnetic density, this gives rise to stable highly super-Chandrasekhar white dwarfs (see e.g. \citealt{dm14,UD3,dm15,Sathya}). Note that a large number of magnetized white dwarfs with surface fields as high as $10^{9}\,$G have been discovered by the Sloan Digital Sky Survey (\citealt{Schmidt1}). It is possible that their central fields are several orders of magnitude larger than their surface fields. To capture the variation of field magnitude $B$, irrespective of the other complicated effects (including the field geometry) that might be involved, we use a profile proposed earlier by \citet{Bandyopadhyay}, modelling $B$ as a function of $\rho$, given by
\begin{equation}
B\left(\frac{\rho}{\rho_{0}}\right) = B_{\rm{s}} + B_{0}\left[1-\rm{exp}{\left(-\eta \left(\frac{\rho}{\rho_{0}}\right)^{\gamma}\right)}\right],
\label{eqn5}
\end{equation} 
where $B_{\rm{s}}$ is the surface magnetic field, $B_{0}$ (similar to the central field) is a parameter with the dimension of $B$. $\eta$ and $\gamma$ are parameters determining how the magnitude of magnetic field decreases from the core to the surface. The magnitude of $\rho_{0}$ is chosen to be about 10 percent of $\rho_{\rm{c}}$, where $\rho_{\rm{c}}$ is the central density. We set $\eta = 0.8$, $\gamma = 0.9$ and $\rho_{0} = 10^{9}\, \rm{g\,cm^{-3}}$ for all our calculations. Close to the surface we have $\rho \rightarrow 0$ and therefore $B \rightarrow B_{\rm{s}}$. This field profile has been used to successfully model neutron stars for quite sometime. Here, with the appropriate change of parameters, we use it for white dwarfs (as was done earlier, \citealt{dm14}). In our simple model we neglect complicated effects such as offset dipoles and magnetic spots which can arise from more complex field structures (see e.g. \citealt{Max,Ven}). Hence, the magnetic field profile can be adequately described by equation (\ref{eqn5}). 

Dividing equations (\ref{eqn1}) and (\ref{eqn2}), we can write
\begin{equation}
\frac{d}{dT}\left(P+P_B\right) = \frac{4ac}{3} \frac{4\pi GM}{L} \frac{T^{3}}{\kappa}.
\label{eqn3}
\end{equation}

\begin{table*}
\begin{center}
\caption{\small $T_{*}$, $\rho_{*}$, and $r_{*}$ for different $L$, when $T_{\rm s}=(L/4\pi R^2 \sigma)^{1/4}$ and $R = 5000\,$km}
\begin{tabular}{cccccccccccccccccccccc}
\hline
\hline
$L/L_{\odot}$ & $T_{*}/\rm{K}$ &  $\rho_{*}/\rm{g\,cm^{-3}}$ & $r_{*}/R$ & $T_{\rm s}/\rm{K}$ \\ \hline 
$10^{-5}$ & $2.332\times10^{6}$ & $1.707\times10^{2}$  & $0.9978$ & $3.847\times10^{3}$ \\ \hline  
$5\times10^{-5}$ & $3.693\times10^{6}$ & $3.403\times10^{2}$  & $0.9965$ & $5.753\times10^{3}$ \\ \hline
$10^{-4}$ & $4.502\times10^{6}$ & $4.580\times10^{2}$ & $0.9958$ & $6.841\times10^{3}$ \\ \hline
$5\times10^{-4}$ & $7.131\times10^{6}$ & $9.129\times10^{2}$  & $0.9933$ & $1.023\times10^{4}$ \\ \hline
$10^{-3}$ & $8.693\times10^{6}$ & $1.229\times10^{3}$ & $0.9918$ & $1.217\times10^{4}$ \\ \hline
$5\times10^{-3}$ & $1.377\times10^{7}$ & $2.449\times10^{3}$  & $0.9871$ & $1.819\times10^{4}$ \\ \hline
$10^{-2}$ & $1.678\times10^{7}$ & $3.296\times10^{3}$ & $0.9844$ & $2.163\times10^{4}$ \\ \hline
\hline
\label{table1}
\end{tabular}
\end{center}
\end{table*}

While the EoS of the matter near the core is that of a non-relativistic degenerate gas, the surface layers have the EoS of a non-degenerate ideal gas. 
At the interface between the degenerate core and the non-degenerate envelope, the density 
($\rho_*$) and temperature ($T_*$) can be related for the non-magnetized case, by equating 
the respective electron pressure on both sides \citep{Shapiro} so that
\begin{equation}
\rho_{*} \approx (2.4\times10^{-8} {\rm{\,g\,cm^{-3}\,K^{-3/2}}})\, \mu_{\rm{e}}T_{*}^{3/2},
\label{eqn4}
\end{equation}
where $\mu_{\rm{e}} \approx 2$ is the mean molecular weight per electron. 
However, in the presence of $B_{\rm{s}}\gtrsim 10^{12}\,$G
(which sometimes is the case in this work) quantum mechanical effects become important and 
equation (\ref{eqn4}) is no longer strictly valid, because the contribution of $\rho_B$ to the density at the interface and its neighbourhood need not be negligible (see e.g. \citealt{heanselbook} for details). After including the quantum mechanical effects, the EoS for the degenerate core depends on the strength of $B$ \citep{VenPot}, while the EoS for the non-degenerate envelope is unaffected. For the non-relativistic electrons, the electron pressure on both sides of the interface can then be equated to give
\begin{eqnarray}
\rho_{*}(B_{*}) = (1.482\times10^{-12} {\rm{\,g\,cm^{-3}\,K^{-1/2}\,G^{-1}}})\, T_{*}^{1/2}B_{*} \nonumber \\ 
\approx (1.482\times10^{-12}{\rm{\,g\,cm^{-3}\,K^{-1/2}\,G^{-1}}})\, T_{*}^{1/2}B_{\rm s}
\label{eqn_rhoT_B}
\end{eqnarray}
as $\rho_{*} \ll \rho_{0}$ (from equation \ref{eqn5}).
The strongly quantizing effects of magnetic fields on the EoS of degenerate white dwarf cores have been studied in detail previously for radially constant field profiles \citep{dm12,dm13}. Although it was found that the interface density for a fixed interface temperature can change by a factor of about 3, owing to the presence 
of the magnetic fields under consideration, the resultant effect on the luminosity of the white dwarf is found to be much more significant, as we discuss in subsequent sections.

For magnetized neutron stars, the cooling rate can be influenced by the suppression of thermal conduction in the direction transverse to the magnetic field lines (see \citealt{hern,Potekhin}). However, it was shown \citep{Tremblay} that unlike neutron stars, changes in conduction rates in white dwarfs do not affect the cooling process because the insulating region is non-degenerate and thermal conduction takes place only in the stellar interior. Moreover, average magnetic fields considered for white dwarfs here are much weaker than those found in neutron stars. Therefore, we choose the core to be isothermal as it is for the non-magnetized white dwarfs. Throughout this paper, we consider white dwarfs with mass $M=M_{\odot}$ which corresponds to radius $R=5000\,$km using Chandrasekhar's relation for white dwarfs (\citealt{Chandra1, Chandra2}). However, the results presented here do not change for other radii (in the range 500 to $5000\,$km) and $M$, unless the surface temperature $T_{\rm{s}}$ is as high as $10^5\,$K.

For non-magnetized white dwarfs ($B = 0$), we substitute $P$ from the EoS of non-degenerate matter (ideal gas), as is in the envelope, and integrate equations (\ref{eqn1}) and (\ref{eqn2}) across the envelope to obtain the $\rho - T$ and $r - T$ profiles. The left- and right-hand panels of Fig. \ref{fig1} show the variations of density and radius, respectively, with temperature in the non-degenerate envelope of a non-magnetized white dwarf, with $T_{\rm{s}}=(L/4\pi R^2 \sigma)^{1/4}$, $\rho(T_{\rm{s}}) = 10^{-10}\, \rm{g\,cm^{-3}}$ and 
$r(T_{\rm{s}}) = R = 5000 \rm{\,km}$, where $\sigma$ is the Stefan-Boltzmann constant. The left-hand panel of Fig. \ref{fig1} shows that the density at a given temperature (and hence given radius) is suppressed with increasing luminosity. We obtain the $r - T$ relations to be straight lines with the same slope for different luminosities, as shown in the right-hand panel of Fig. \ref{fig1}.
Once we obtain $\rho-T$ and $r-T$ profiles for the given boundary conditions, we can find $T_{*}$ and $\rho_{*}$ 
by solving for the $\rho-T$ profile along with equation (\ref{eqn4}) as shown in the left-hand panel of Fig. \ref{fig1}. This works because the $\rho-T$ profile is valid in the whole envelope whereas 
equation (\ref{eqn4}) is valid only at the interface. Once we know $T_{*}$, we can also find $r_{*}$ from the $r-T$ 
profile with the right-hand panel of Fig. \ref{fig1}. 
Because $T_{*}$ is different for different luminosities, the corresponding $T-r$ lines should 
originate from different temperatures at the interface.

\begin{figure*}
  \begin{subfigure}[tp]{0.5\linewidth}
    \centering
    \includegraphics[width=0.97\linewidth]{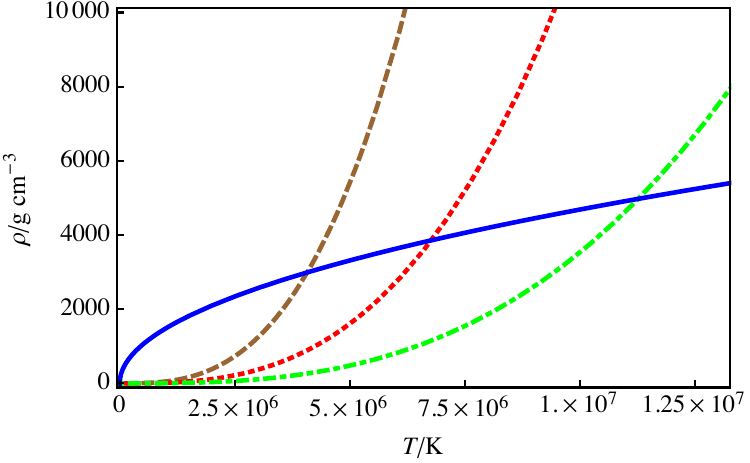}
  \end{subfigure}
  \begin{subfigure}[tp]{0.5\linewidth}
    \centering
    \includegraphics[width=0.93\linewidth]{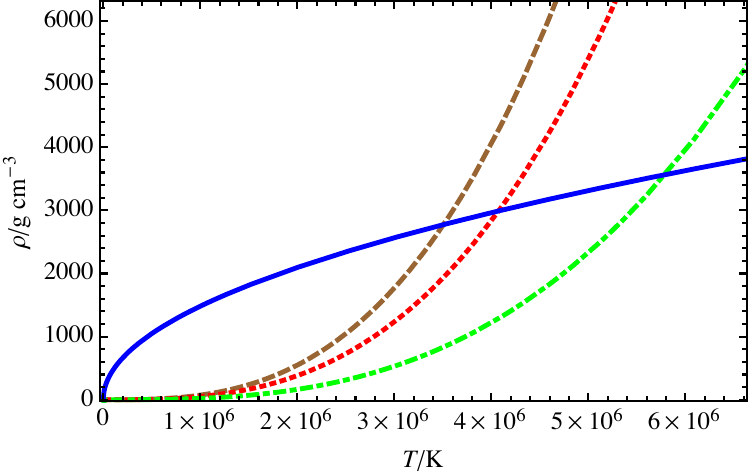}
  \end{subfigure}  
  \caption{
 {\it Left-hand panel:} variation of density with temperature for $B\equiv (B_{\rm{s}},B_0) = (10^{12}\,\rm{ G},10^{14}\,\rm{G})$ and different $L$: $10^{-5}\, L_{\odot}$ (dashed line), $10^{-4}\, L_{\odot}$ (dotted line) and $10^{-3}\, L_{\odot}$ (dot-dashed line).  $\rho_{*}$ and $T_{*}$ are obtained from the intersection of the $\rho-T$ profiles with equation (\ref{eqn_rhoT_B}) (solid line).
 {\it Right-hand panel:} variation of density with temperature for $L = 10^{-5}\, L_{\odot}$ and different $B$: $(10^{12}\,\rm{G}, 5\times10^{13}\,\rm{G})$ (dashed line), $(10^{12}\,\rm{G}, 10^{14}\,\rm{G})$ (dotted line) and $(10^{12}\,\rm{G}, 5\times10^{14}\,\rm{G})$ (dot-dashed line). The $\rho_{*}$ and $T_{*}$ are obtained from the intersection of the $\rho-T$ profiles with equation (\ref{eqn_rhoT_B}) (solid line).
 }
  \label{fig2} 
\end{figure*}

Table \ref{table1} shows the variation of $T_{*}$, $\rho_{*}$ and $r_{*}$ as $L$ changes 
in the range $10^{-5}L_\odot\leq L\leq 10^{-2}L_\odot$, for given $T_{\rm s}$ and $R$ of non-magnetic white dwarfs. We see that, as $L$ increases, $T_{*}$ and $\rho_*$ increase whereas $r_{*}$ decreases. Hence, as the luminosity of a non-magnetized white dwarf increases, the interface shifts inwards and the degenerate region shrinks in volume. However, for the observed range of luminosities, the decrease in volume of the degenerate region is quite small. Also $\left|\Delta T/\Delta r\right| = \left|(T_{\rm s} - T_*)/(R - r_*)\right|$ does not vary appreciably with luminosity and is almost constant.

Now we consider $B\neq 0$ and vary both $B_{0}$ and $B_{\rm{s}}$ to find the temperature profile for a radially varying field. Here, we consider $B$ to be only varying with density and white dwarfs to be approximately spherically symmetric. 
It is generally believed that the magnetic field strength at the surface of a white dwarf is several orders of magnitude smaller than the central field strength (see, e.g., \citealt{Fujisawa,dm14,Sathya}). This is mainly because of 
the consideration of the field to be fossil field of the original star which is 
expected to have a stronger field in the core than its surface in addition to dynamo effects that can replenish and make the core field stronger (see, however, 
\citealt{potter}). Therefore, we consider a realistic density dependent magnetic field profile such that the magnetic field strength decreases from the core of the white dwarf to its surface. We choose $10^{-5} L_{\odot} \le L \le 10^{-2} L_{\odot}$, as for the $B=0$ case, and vary the magnitudes of $B_{\rm{s}}$ and $B_{0}$, keeping $\eta$ and $\gamma$ constant, to investigate how $T_{*}$, $r_{*}$, and the temperature profile change. 
It is important to choose the central and surface fields (and hence 
corresponding $B_0$ and $B_{\rm{s}}$ in equation \ref{eqn5}) keeping stability 
criteria in mind. It was argued earlier \citep{brait}
that the magnetic energy should be well below the gravitational energy
in order to form a stable white dwarf and following that criterion we simulated
highly magnetized stable white dwarfs \citep{dm15,Sathya}. 
In this work, we explore white dwarfs with central and surface fields 
that give rise to stable configurations as described earlier \citep{dm15,Sathya}.
However, for simplicity, here we also fix radius ($R=5000\,$km)
throughout even though this need not be the case for all chosen fields.
Realistically, all chosen sets of $B_{\rm{s}}$ and $B_0$ lead to stable stars with 
different corresponding $R$. Nevertheless,
in this work, $R$ does not play any significant role (except
to compute $T_{\rm{s}}$) and a slight change in $R$ with the change
in fields does not alter our main conclusion. Hence, we keep them fixed.
In addition, we also discuss a (hypothetical) case with constant $B$ for 
completeness, restricting the field in order to equilibrate the star at 
$R=5000\,$km.

We are interested in the surface layers that are non-degenerate, so we can substitute $P$ in terms of $\rho$ in equation (\ref{eqn3}) by the ideal gas EoS, as for the $B=0$ case, to obtain
\begin{equation}
\frac{d}{dT}\left(\frac{\rho k_BT}{\mu m_{\mu}} + \frac{B^2}{8\pi}\right) = \frac{4ac}{3} \frac{4\pi GM}{L} \frac{T^{3}}{\kappa_{B}}.
\label{eqn6}
\end{equation}
and thence
\begin{eqnarray}
(5.938\times10^{7} {\rm{cm^{2}\,s^{-2}\,K^{-1}}})\, \rho + (5.938\times10^{7} {\rm{cm^{2}\,s^{-2}\,K^{-1}}})\, T \frac{d\rho}{dT}\nonumber \\ 
+ 0.0796 B \frac{dB}{d\rho} \frac{d\rho}{dT} =  \frac{(9.218\times10^{-9} {\rm{g^{2}\,cm^{-1}\,s^{-3}\,K^{-5.5}}})}{L} \frac{T^{4.5}}{\rho}B^2.\nonumber\\
\label{eqn7}
\end{eqnarray}
From equation (\ref{eqn2}), we have
\begin{equation}
\frac{dr}{dT} = -(6.910\times10^{-35} {\rm{g^{2}\,cm^{-4}\,s^{-1}\,K^{-5.5}}}) \frac{T^{4.5}B^2}{\rho\left(\rho + \frac{B^2}{2.261\times10^{22}}\right)} \frac{r^{2}}{L}.
\label{eqn8}
\end{equation}

As for the $B=0$ case, equations (\ref{eqn7}) and (\ref{eqn8}) are simultaneously solved with boundary 
conditions at the 
surface: $\rho(T_{\rm{s}}) = 10^{-10}\, \rm{g\,cm^{-3}}$ and $r(T_{\rm{s}}) = R = 5000\,$km. 
As before, once we obtain the $\rho-T$ and $r-T$ profiles for the given boundary conditions, we can find 
$T_{*}$ and $\rho_{*}$ 
by solving for the $\rho-T$ profile along with equation (\ref{eqn_rhoT_B}), as shown in Fig. \ref{fig2}. 
Once we know $T_{*}$, we can also find $r_{*}$ from the $r-T$ profile.

\begin{figure*}
  \begin{subfigure}[tp]{0.5\linewidth}
    \centering
    \includegraphics[width=0.97\linewidth]{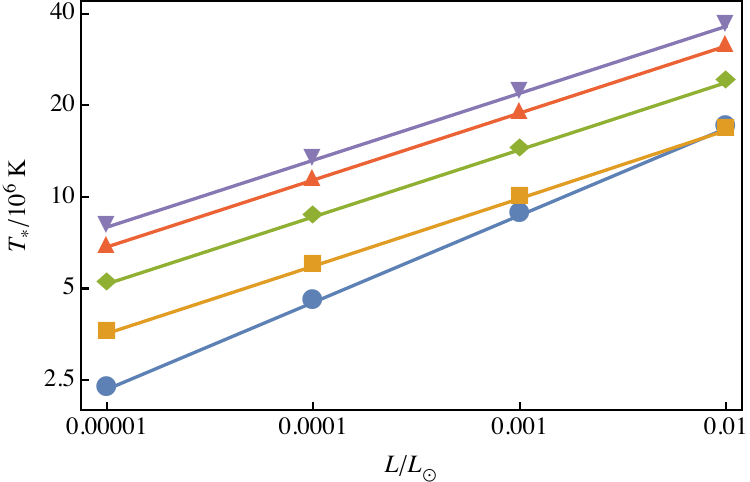}
  \end{subfigure}
  \begin{subfigure}[tp]{0.5\linewidth}
    \centering
    \includegraphics[width=0.97\linewidth]{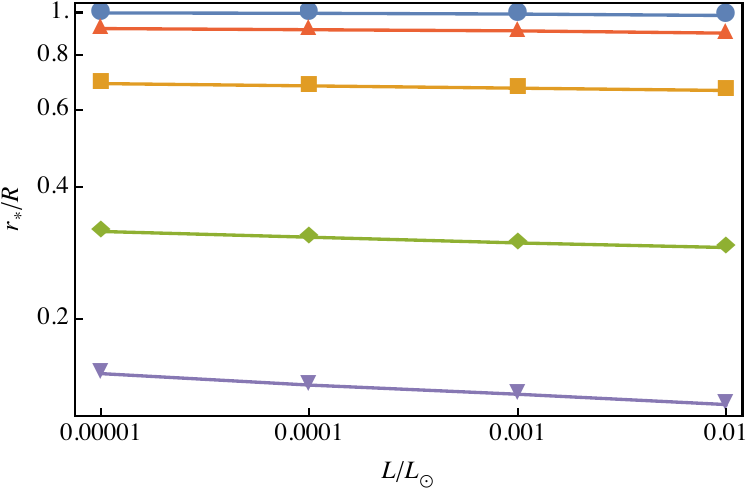}
  \end{subfigure}  
  \caption{
 {\it Left-hand panel:} variation of temperature at interface with luminosity for different $B$: $(0\,\rm{G}, 0\,\rm{G})$ (circles), $(5\times10^{11}\,\rm{G}, 10^{14}\,\rm{G})$ (squares), $(10^{12}\,\rm{G}, 3\times10^{14}\,\rm{G})$ (diamonds), $(3\times 10^{12}\,\rm{G},4\times 10^{14}\,\rm{G})$ (upward triangles) and $(5\times 10^{12}\,\rm{G}, 5\times10^{14}\,\rm{G})$ (downward triangles).
 {\it Right-hand panel:} variation of radius at interface with luminosity for different $B$: $(0\,\rm{G}, 0\,\rm{G})$ (circles), $(10^{11}\,\rm{G}, 5\times10^{14}\,\rm{G})$ (squares), $(10^{12}\,\rm{G}, 5\times10^{14}\,\rm{G})$ (diamonds), $(7\times10^{12}\,\rm{G}, 0\,\rm{G})$ (upward triangles) and $(5\times 10^{12}\,\rm{G}, 5\times10^{14}\,\rm{G})$ (downward triangles). 
 }
  \label{fig3} 
\end{figure*}


In the left- and right-hand panels of Fig. \ref{fig3}, we show the variation of $T_{*}$ and $r_{*}$ respectively, for different $B\equiv(B_{\rm{s}},B_0)$ and 
$L$. Note that here the point of computation is interface radius and hence the luminosity is actually of 
interface radius ($L_*$). However this $L_*$ is effectively the same as $L$ (hence we use them 
interchangeably). 
From the left-hand panel of Fig. \ref{fig3}, we see that $T_{*}$ increases with increasing $B_{\rm{s}}$, $B_{0}$, and $L$. For a given $(B_{\rm{s}}, B_{0})$, $T_{*}$ increases as $L$ increases. However, the fractional change in $T_{*}$ with the change in $L$ decreases as $B_{\rm{s}}$ and $B_{0}$ increase. In other words, the increase of $T_{*}$ owing to the increase of 
$L$, is somewhat saturated by the increase in $B$. 
For a fixed $L$, $T_{*}$ increases considerably with $B$ only when $B_{\rm{s}} \ge 5\times10^{11}\,$G and
$B_{0} \ge 10^{14}\,$G. For a constant $B$, the change in $T_*$ at a given $L$ is  
very small compared to that in nonmagnetized case.
Also, for a given set of $B_{\rm{s}}$ and $B_{0}$, $r_{*}$ decreases with $L$, as seen in the right-hand panel of Fig. \ref{fig3}. Therefore, the interface moves inwards with an increase in $L$ for a given $(B_{\rm{s}}, B_{0})$. However, unless $B$ is very high, the change in $r_*$ is not significant.
The radius $r_{*}$ decreases with the increase of $B$, with the change being considerable for $B_{\rm{s}} \ge 10^{10}\,$G and $B_{0} \ge 10^{14}\,$G. Therefore, the interface moves inwards with an
increase of magnetic field strength and an increase of luminosity. 
The right-hand panel of Fig. \ref{fig3} also includes the result for the (hypothetical) case of constant $B = 7\times10^{12}\,$G throughout the star. Interestingly, this shows the same trend as varying $B$, with a very small change in $r_*$.

\begin{figure*}
  \begin{subfigure}[tp]{0.5\linewidth}
    \centering
    \includegraphics[width=0.97\linewidth]{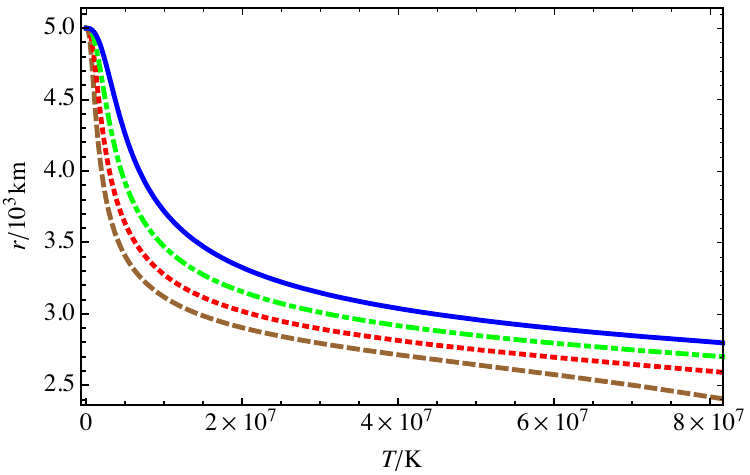}
  \end{subfigure}
  \begin{subfigure}[tp]{0.5\linewidth}
    \centering
    \includegraphics[width=0.97\linewidth]{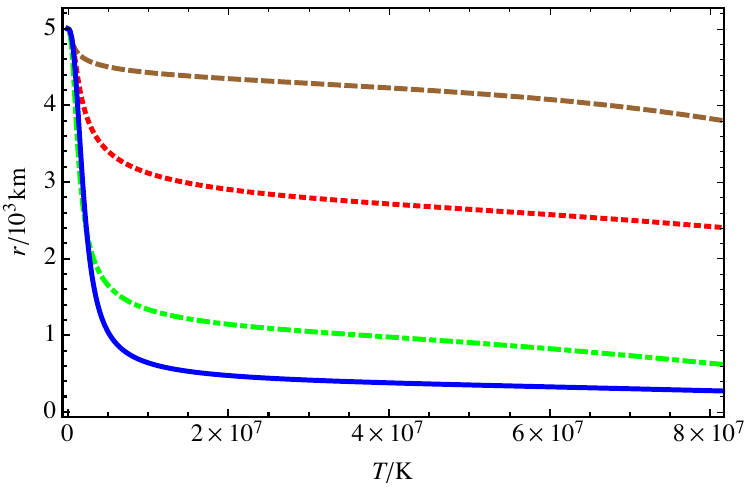}
  \end{subfigure}  
  \caption{
 {\it Left-hand panel:} variation of radius with temperature for $B = (10^{12}\,\rm{G}, 10^{14}\,\rm{G})$ and different luminosities: $10^{-5}\, L_{\odot}$ (dashed line), $10^{-4}\, L_{\odot}$ (dotted line), $10^{-3}\, L_{\odot}$ (dot-dashed line) and $10^{-2}\, L_{\odot}$ (solid line).
 {\it Right-hand panel:} variation of radius with temperature for $L = 10^{-5}\, L_{\odot}$ and different magnetic fields: $ (10^{11}\, \rm{G}, 10^{14}\, \rm{G})$ (dashed line), $ (10^{12}\, \rm{G}, 10^{14}\, \rm{G})$ (dotted line), $ (10^{12}\, \rm{G}, 5\times10^{14}\, \rm{G})$ (dot-dashed line) and $ (5\times 10^{12}\,\rm{G}, 5\times10^{14}\, \rm{G})$ (solid line).
 }
  \label{fig4} 
\end{figure*}

As shown in Fig. \ref{fig4}, unlike for the non-magnetized white dwarf case, the $r - T$ profile is no longer linear for any $L$. Also, as $L$ increases, $dT/dr$ near the surface increases.
The gradient $dT/dr$ near the surface decreases with the increase in magnitude of $B$. 
Therefore, the temperature-fall rate near the surface increases with luminosity and decreases with field strength.
The density $\rho_*$ also increases, like the $B=0$ case, with the increase of $L$ or $B$, as $\rho_* \propto T_*^{1/2}B$ from equation (\ref{eqn_rhoT_B}).



\section{Variation of luminosity with magnetic field}
\label{sec3}
In this section, we determine how the luminosity of a white dwarf changes as the magnetic field strength increases such that
\vspace{0.05 in}
\\(i) the interface radius for a magnetized white dwarf is the same as that for a non-magnetized white dwarf, $r_{*,B\neq0}=r_{*,B=0}$, and
\\(ii) the interface temperature for a magnetized white dwarf is the same as that for a non-magnetized white dwarf, 
$T_{*,B\neq0}=T_{*,B=0}$. 
\vspace{0.1in} 

The motivation for fixing $r_*$ or $T_*$ between non-magnetized and magnetized cases is to better constrain the individual components (gravitational, thermal and magnetic) of the conserved total energy of the magnetized white dwarf. 
For the fixed $r_{*}$ case, we assume that the increase in magnetic field energy is compensated by an equal decrease in the thermal energy of the isothermal electron-degenerate white dwarf core while the gravitational potential energy remains unaffected (owing to fixed $r_{*}$ and $R$). This is justified by the decrease in $T_{*}$ (and therefore $L$) with increase in $B$ (see Table \ref{table5}). 

For the fixed $T_{*}$ case, we assume that the increase in magnetic field energy is compensated by an equal decrease in gravitational potential energy of the white dwarf whereas the thermal energy is unchanged (owing to fixed core temperature $T_{\rm core}=T_{*}$). This indeed makes sense because, with increase in $B$ for fixed $T_{*}$, $r_{*}$ decreases (see Table \ref{table6}) with more and more electron-degenerate mass concentrated near the centre of the white dwarf, thereby reducing the effective gravitational potential energy.
Indeed, observationally, it was found (\citealt{ferra}) that the temperature of white 
dwarfs does not vary much with magnetic field, although the maximum 
$B_{\rm{s}}$ observed so far is $B \lesssim 10^9\,$G, which is quite small compared to the 
fields considered here.

\begin{table*}
\begin{center}
\caption{\small Variation of luminosity with magnetic field for fixed $r_{*}=0.9978\,R$}
\begin{tabular}{cccccccccccccccccccccc}
\hline
\hline
\centering
$B/\rm{G}=(B_{\rm{s}}/G,B_0/G)\,$ & $L/L_{\odot}$ &  $T_{*}/\rm{K}$ & $\rho_{*}/\rm{g\,cm^{-3}}$ & $T_{\rm{s}}/\rm{K}$ \\ \hline
$(0, 0)$                 & $1.00\times10^{-5}$ & $2.332\times10^{6}$ & $1.707\times10^{2}$ & $3.85\times10^{3}$ \\ \hline
$(10^9,6\times10^{13})$ & $2.53\times10^{-7}$ & $4.901\times10^{5}$ & $1.037\times10^{0}$ &$1.53\times10^{3}$ \\ \hline
$(2\times10^{9},4\times10^{13})$ & $2.07\times10^{-8}$ & $2.737\times10^{5}$ & $1.551\times10^{0}$ &$8.21\times10^{2}$ \\ \hline
$(5\times10^{9},2\times10^{13})$ & $3.96\times10^{-8}$ & $3.262\times10^{5}$ & $4.232\times10^{0}$ &$9.65\times10^{2}$ \\ \hline
$(10^{10},10^{13})$ & $1.02\times10^{-6}$ & $7.189\times10^{5}$ & $1.257\times10^{1}$ &$2.17\times10^{3}$ \\ \hline
$(2\times10^{10},6\times10^{12})$ & $1.22\times10^{-6}$ & $7.616\times10^{5}$ & $2.587\times10^{1}$ &$2.27\times10^{3}$ \\ \hline
$(2\times10^{10},8\times10^{12})$ & $4.40\times10^{-9}$ & $2.063\times10^{5}$ & $1.346\times10^{1}$ &$5.57\times10^{2}$ \\ \hline
$(5\times10^{10},4\times10^{12})$ & $2.59\times10^{-8}$ & $3.185\times10^{5}$ & $4.182\times10^{1}$ &$8.68\times10^{2}$ \\ \hline
$(10^{11},2\times10^{12})$ & $1.09\times10^{-6}$ & $7.721\times10^{5}$ & $1.302\times10^{2}$ &$2.21\times10^{3}$ \\ \hline
$(5\times10^{11},10^{12})$ & $2.93\times10^{-9}$ & $2.206\times10^{5}$ & $3.480\times10^{2}$ &$5.03\times10^{2}$ \\ \hline
\hline
\label{table5}
\end{tabular}
\end{center}
\end{table*}

\begin{table*}
\begin{center}
\caption{\small Variation of luminosity with magnetic field for fixed $T_{*}=2.332\times10^{6}\,\rm{K}$}
\begin{tabular}{cccccccccccccccccccccc}
\hline
\hline
\centering
$B/\rm{G}=(B_{\rm{s}}/G,B_0/G)\,$ & $L/L_{\odot}$ & $\rho_{*}/\rm{g\,cm^{-3}}$ & $r_{*}/R$ & $T_{\rm{s}}/\rm{K}$\\ \hline
$(0, 0)$ & $1.00\times10^{-5}$ & $1.707\times10^{2}$ & $0.9978$ & $3.85\times10^{3}$\\ \hline
$(10^{11}, 5\times10^{14})$ & $1.26\times10^{-6}$ & $2.263\times10^{2}$ & $0.6910$ & $2.29\times10^{3}$\\ \hline
$(2\times10^{11}, 5\times10^{14})$ & $6.77\times10^{-7}$ & $4.526\times10^{2}$ & $0.5830$ & $1.96\times10^{3}$\\ \hline
$(5\times10^{11}, 5\times10^{14})$ & $2.98\times10^{-7}$ & $1.132\times10^{3}$ & $0.4342$ & $1.60\times10^{3}$\\ \hline
$(10^{12}, 10^{14})$ & $7.93\times10^{-7}$ & $2.263\times10^{3}$ & $0.7131$ & $2.04\times10^{3}$\\ \hline
$(10^{12}, 5\times10^{14})$ & $1.60\times10^{-7}$ & $2.263\times10^{3}$ & $0.3326$ & $1.37\times10^{3}$\\ \hline
$(2\times10^{12}, 10^{14})$ & $4.26\times10^{-7}$ & $4.526\times10^{3}$ & $0.6236$ & $1.75\times10^{3}$\\ \hline
$(2\times10^{12}, 5\times10^{14})$ & $8.57\times10^{-8}$ & $4.526\times10^{3}$ & $0.2491$ & $1.17\times10^{3}$\\ \hline
$(5\times10^{12}, 10^{14})$ & $1.87\times10^{-7}$ & $1.132\times10^{4}$ & $0.5055$ & $1.42\times10^{3}$\\ \hline
$(5\times10^{12}, 5\times10^{14})$ & $3.76\times10^{-8}$ & $1.132\times10^{4}$ & $0.1698$ & $9.52\times10^{2}$\\ \hline
\hline
\label{table6}
\end{tabular}
\end{center}
\end{table*}

We calculate $L$ for various magnetic field profiles, such that either $r_{*}$ or $T_{*}$ is the same as for the non-magnetized white dwarf with $L=10^{-5} L_{\odot}$. Overall, it turns out
that, 
depending on the field strength and profile, 
the magnetic fields have a significant impact on the equilibrium stellar structure.

Note importantly that 
$B\lesssim 10^9\,$G practically has no effect on 
the white dwarf mass-radius relation as long as it is assumed
to be constant throughout the star. However, a white dwarf
with a surface field $B_{\rm{s}} \approx 10^9\,$G (which we could observe) can have a much stronger central field (up to $B_{\rm{s}} \approx 10^{14}\,$G). 
This could lead to massive, even super-Chandrasekhar, white dwarfs, depending on the field profiles 
\citep{dm15,Sathya}. Nevertheless, here, we assume a fixed  
initial mass and radius for the white dwarfs of a fixed age. 
This is possible for appropriate choice of field profiles
along with the chosen respective central and surface fields.

\subsection{Fixed interface radius}
\label{sec3.1}
We assume a magnetic field profile as given by equation (\ref{eqn5}) and find the variation of luminosity with a change in $B_{\rm{s}}$ and $B_{0}$ so that the interface radius is same as for the non-magnetic case. Note that for $B=0$ and $L=10^{-5}\, L_{\odot}$, we have found $r_{*}=0.9978\, R$, $\rho_{*}=170.7 \,\rm{g\,cm^{-3}}$ and $T_{*} = 2.332\times10^{6}\,$K (Table 1). We solve equations (\ref{eqn7}) and (\ref{eqn8}) using the same boundary conditions as in section \ref{sec2} but this time vary $L$ in order to fix $r_{*}=0.9978\,R$.

Interestingly, Table \ref{table5} shows that $L$ and $T_{*}$ both decrease as the magnetic field strength increases. However, the change is appreciable only for $B_{\rm{s}} \ge 10^{10}\,$G or $B_{0} \ge 10^{13}\,$G with  
$L$ becoming quite low $L \approx 10^{-6}\,L_{\odot}$, and lower for white dwarfs with $(B_{\rm{s}},\,B_{0}) = (2\times10^{10}\,\rm{G},\,7\times10^{12}\,\rm{G})$ and higher. This can make it difficult to detect such highly magnetized white dwarfs.

Motivated by the high $B$ cases in the right-hand panel of Fig. \ref{fig3}, if $r_*$ 
is chosen to be smaller than its non-magnetized counterpart (for a given $T_*$), $T_{\rm{s}}$ and $L$ also decrease more compared to the
non-magnetic case for a fixed radius of the star, because then $T$ can decrease more (over a larger region) from the interface to the surface.

\subsection{Fixed interface temperature}
\label{sec3.2}
Here, we solve equations (\ref{eqn7}) and (\ref{eqn8}) as in section \ref{sec2}, but this time we vary $L$ to get $T_{*} = 2.332\times10^{6}\,$K, using the same boundary conditions as in section \ref{sec2}. We find that $L$ has to decrease as $B$ increases for $T_{*}$ to be unchanged. From Table \ref{table6}, we see that $L$ becomes very small when $B_{\rm{s}} > 2\times10^{11}$ and $B_{0} \ge 2\times10^{14}\,$G. We also see that $r_{*}$ decreases with increase in magnetic field strength.
However, with a higher $T_*$, $T_{\rm{s}}$ and $L$ could still be lower as $B$ 
increases, if we relax the assumption of fixed radius for the white dwarf and consider
it to be increased, as is the case in the presence of toroidally
dominated fields (see, e.g., \citealt{dm15,Sathya}).

\section{Cooling in the presence of a magnetic field and post cooling temperature profile}
\label{sec4}
In this section, we discuss briefly how the cooling time-scale of a non-magnetized white dwarf can be evaluated when we know the $L-T$ relation. 
Motivated by the analysis of the cooling evolution for non-magnetized white dwarfs, we estimate $L-T$ relations for the magnetic cases in section \ref{sec3} by fitting power laws of the form $L = \alpha T^{\gamma}$ for different field strengths.
Using those $L-T$ relations, we implement cooling over time to find the present interface temperature, $T_{*, \rm{pr}}$, from the initial interface temperature $T_{*, \rm{in}}$ for $\tau = 10\,$Gyr.

\subsection{Cooling time-scale for white dwarfs}
\label{sec4.1}
Here, we briefly recapitulate the discussion of white dwarf cooling rate \citep{Mestel,Schwarzschild}. Then, we discuss the effect of magnetic field on the specific heat and the cooling evolution of white dwarfs.

\subsubsection{Non-magnetized white dwarfs}
\label{sec4.1.1}
The thermal energy of the ions is the only significant source of energy that can be radiated when a star enters the white dwarf stage because most of the electrons occupy the lowest energy states in a degenerate gas. Also, the energy release from neutrino emission is considerable only in the very early phase when the temperature is high. 

The thermal energy of the ions and the rate at which it is transported to the surface to be radiated depends on the specific heat, which in turn depends significantly on the physical state of the ions in the core. The cooling rate of a white dwarf $-dU/dt$ can be equated to $L$ to give (\citealt{Shapiro})
\begin{equation}
L=-\frac{d}{dt} \int c_{\rm{v}} dT = (2\times10^{6}\, {\rm erg\, s^{-1}\, K^{-7/2}}) \frac{Am_{\mu}}{M_{\odot}} T^{7/2},
\label{eqnx}
\end{equation}
where $c_{\rm{v}}$ is the specific heat at constant volume and $A$ is the atomic weight.

For $T \gg T_{\rm{g}}$ (where $T_{\rm{g}}$ corresponds to a point at which the ion kinetic energy exceeds
its vibrational energy), $c_{\rm v} \approx 3k_{\rm{B}}/2$, where $k_{\rm{B}}$ is Boltzmann constant. This gives us
\begin{eqnarray}
\left(T^{-5/2}-{T_{0}}^{-5/2}\right) = (3.3\times10^{6}\, {\rm erg\, s^{-1}\, K^{-7/2}}) \frac{Am_{\mu}}{M_{\odot}} \frac{(t-t_{0})}{k_{\rm{B}}} \nonumber \\
= (2.4058\times10^{-34}\, {\rm s^{-1}\,K^{-5/2}})\tau,
\label{eqn9}
\end{eqnarray}
where $T_{0}$ is the initial temperature (before cooling starts), $T$ is the present temperature at time $t$ and $\tau = t-t_{0}$ is the age of the white dwarf. Using equations (\ref{eqnx}) and (\ref{eqn9}), 
we can find $T$ at the interface and $L$ for various $\tau$ which corresponds to the present age of the white dwarf. We calculate $T$ for $T_{*}=T_{0}$ given in Table \ref{table1} and $\tau = 10\,$Gyr = $3.1536\times10^{17}\,$s. It is important to note that $\tau$ cannot exceed $13.8\,$Gyr, which is the present age of the Universe. 

\begin{figure*}
  \begin{subfigure}[tp]{0.5\linewidth}
    \centering
    \includegraphics[width=0.97\linewidth]{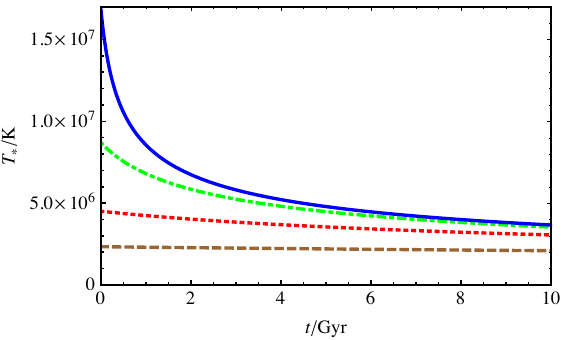}
  \end{subfigure}
  \begin{subfigure}[tp]{0.5\linewidth}
    \centering
    \includegraphics[width=0.93\linewidth]{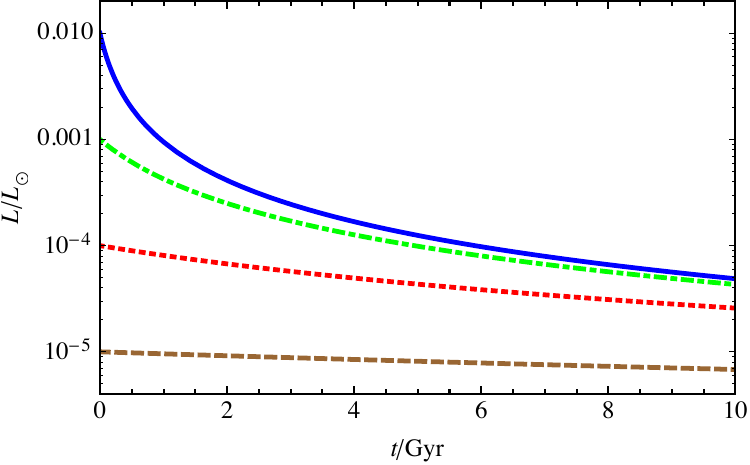}
  \end{subfigure}  
  \caption{
 {\it Left-hand panel:} variation of interface temperature with time for a 
non-magnetized white dwarf with different initial luminosities: $10^{-5} L_{\odot}$ (dashed line), $10^{-4} L_{\odot}$ (dotted line), $10^{-3} L_{\odot}$ (dot-dashed line) and $10^{-2} L_{\odot}$ (solid line).
 {\it Right-hand panel:} variation of luminosity with time for a non-magnetized
 white dwarf with different initial luminosities: $10^{-5} L_{\odot}$ (dashed line), $10^{-4} L_{\odot}$ (dotted line), $10^{-3} L_{\odot}$ (dot-dashed line) and $10^{-2} L_{\odot}$ (solid line).
 }
  \label{fig5} 
\end{figure*}

From the left-hand panel of Fig. \ref{fig5}, it can be seen that cooling at the interface is considerable only for higher luminosities ($L\ge 10^{-3} L_{\odot}$) and that white dwarfs spend most of the time near their present temperature. This is why we have retained the terms associated with $T_0$ in above expressions. From the right-hand panel of Fig. \ref{fig5}, it can be seen that even after $10\,$Gyr, $L$ decreases only by 1 order of magnitude, which explains why many white dwarfs have not faded from view, even though their initial luminosities may have been quite low.

Convection might also result in shorter cooling time-scales owing to more efficient energy transfer but it has been shown not to be significant \citep{LambHorn,FonHorn} to a first-order approximation. This is because convection does not influence the cooling time until the base of the convection zone reaches the degenerate reservoir of thermal energy and couples the surface with the reservoir. This occurs for surface temperatures much lower than what we have considered here. It was also shown by \citet{Tremblay} that convective energy transfer is significantly hampered when the magnetic pressure dominates over the thermal pressure. Note that, although we have assumed simple self-similarity of the cooling process
up to the age of $10\,$Gyr, a more accurate calculation of the
cooling of non-magnetic white dwarfs reveals that it is not
strictly the case \citep{Hansen}. However, this choice is justified by the simple and exploratory nature of our study.

\subsubsection{Specific heat and cooling rate in the presence of magnetic field}
\label{sec4.1.2}
A magnetic field can, in principle, affect the state of the ionic core and thus its thermodynamic properties, such as the specific heat. The relevant parameter to quantify this effect is 
\begin{equation}
b = \frac{\omega_{B}}{\omega_{p}},
\end{equation}
where 
\begin{equation}
\omega_{B} = \frac{ZeB}{Mc}, \,\,\,{\rm and}\,\,\, \omega_{p} = \sqrt{\frac{4\pi Z^{2}e^{2}n}{M}},
\end{equation}
are the ion cyclotron and ion plasma frequencies, respectively. Here $n$ is the number density of the ions,
$e$ is the electric charge and $\omega_p$ is the effective Debye frequency of the ionic lattice. We would expect the effect of the magnetic field on the ionic core to be strong when $b \ge 1$, when the cyclotron frequency is comparable to or larger than the Debye frequency of the lattice. 

The effect of magnetic fields on a Body Centered Cubic (BCC) Coulomb lattice was studied 
by \citet{Baiko} and it was concluded that there is an appreciable change of the specific heat only for $b \gg 1$ except when $T \ll \theta_D$ (Debye temperature). For almost all the white dwarfs that we consider $B < 10^{12}\,$G at the interface. This corresponds to $b \le 1$. Furthermore, the interface temperature is not significantly smaller than $\theta_D$. So, we are justified in working with a specific heat appropriate for a non-magnetized system despite the presence of a magnetic field. 

In the future, it will be of interest to study the effect of much stronger magnetic fields on the ionic core and its specific heat. In particular, if the magnetic field is strong enough to cause Landau quantization of the electron gas in the core, it could change the effective ion-ion interaction as mediated by the electrons. This would be in addition to the direct effect of the field on the ionic core described above. The effect of a magnetic field on the phonon spectrum of ions in conventional solid state systems has been investigated and found to be weak for field strengths appropriate to these systems \citep{Holz}. However, the effect might be appreciable if fields of the order of $10^{15}\,$G arise and could result in very interesting physics. 

\subsection{Fixed interface radius} 
\label{sec4.2}

We find the $L =\alpha T^{\gamma}$ relations for different $B$ from section \ref{sec2} (see the left-hand panel of Fig. \ref{fig3} shown for interface). From Table \ref{table5}, we also know the initial interface luminosity at the onset of cooling, $L_{*,\rm{in}}$ (the 
luminosity computed at $r_*$), and the corresponding initial interface temperature, $T_{*,\rm{in}}$, for different field strengths. Using these in the cooling evolution (equation \ref{eqnx}), we calculate 
the present interface temperature, $T_{*,\rm{pr}}$, for different 
$B$ and $r_{*}=0.9978\, R$, as given in Table \ref{table7}.

\begin{table*}
\begin{center}
\caption{\small Change in $T_{*}$ with time due to the presence of a magnetic field for fixed $r_{*}=0.9978\,R$}
\begin{tabular}{cccccccccccccccccccccc}
\hline
\hline
\centering
$B/\rm{G}=(B_{\rm{s}}/\rm{G},B_0/\rm{G})\,$ & $T_{*,\rm{in}}/\rm{K}\,$ & $L_{\rm{in}}/L_{\odot}$ & $L(T)/{\rm erg\ s^{-1}}$ & $T_{*,\rm{pr}}/\rm{K}$ \\ \hline
$(0, 0)$                 & $2.332\times10^{6}$ & $1.00\times10^{-5}$ & $2.013\times10^{6}T^{3.500}$ & $2.223\times10^{6}$ \\ \hline
$(10^{9},6\times10^{13})$ & $4.901\times10^{5}$ & $2.53\times10^{-7}$ & $2.288\times10^{4}T^{3.971}$ & $4.874\times10^{5}$ \\ \hline
$(2\times10^{9},4\times10^{13})$ & $2.737\times10^{5}$ & $2.07\times10^{-8}$ & $1.551\times10^{3}T^{4.172}$ & $2.735\times10^{5}$ \\ \hline
$(5\times10^{9},2\times10^{13})$ & $3.262\times10^{5}$ & $3.96\times10^{-8}$ & $1.665\times10^{3}T^{4.160}$ & $3.258\times10^{5}$ \\ \hline
$(10^{10},10^{13})$ & $7.189\times10^{5}$ & $1.02\times10^{-6}$ & $2.951\times10^{4}T^{3.943}$ & $7.081\times10^{5}$ \\ \hline
$(2\times10^{10},6\times10^{12})$ & $7.616\times10^{5}$ & $1.22\times10^{-6}$ & $2.474\times10^{4}T^{3.952}$ & $7.488\times10^{5}$ \\ \hline
$(2\times10^{10},8\times10^{12})$ & $2.063\times10^{5}$ & $4.40\times10^{-9}$ & $1.627\times10^{2}T^{4.328}$ & $2.062\times10^{5}$ \\ \hline
$(5\times10^{10},4\times10^{12})$ & $3.185\times10^{5}$ & $2.59\times10^{-8}$ & $3.277\times10^{2}T^{4.263}$ & $3.182\times10^{5}$ \\ \hline
$(10^{11},2\times10^{12})$ & $7.721\times10^{5}$ & $1.09\times10^{-6}$ & $7.099\times10^{3}T^{4.032}$ & $7.606\times10^{5}$ \\ \hline

$(5\times10^{11},10^{12})$ & $2.206\times10^{5}$ & $2.93\times10^{-9}$ & $2.407\times10^{1}T^{4.428}$ & $2.206\times10^{5}$ \\ \hline
\hline
\label{table7}
\end{tabular}
\end{center}
\end{table*}
We find that $L$ decreases with increasing $B$. With the increase of field strength, the coefficient $\alpha$ in the $L=\alpha T^{\gamma}$ relation decreases whereas the exponent $\gamma$ increases. Moreover, increasing $B$ results in slower cooling of the white dwarf.

\subsection{Fixed interface temperature}
\label{sec4.3}
As above, the $L=\alpha T^{\gamma}$ relations for different $B$ are obtained from section \ref{sec2} (see the left-hand panel of Fig. \ref{fig3}) and $L_{*,\rm{in}}$ for different fields are obtained from Table \ref{table6}. We then calculate $T_{*,\rm{pr}}$ for the different $B$ and $T_{*} = 2.332\times10^{6}\,$K using equation (\ref{eqnx}), as given in Table \ref{table8}.

\begin{table*}
\begin{center}
\caption{\small Change in $T_{*}$ with time due to the presence of a magnetic field for fixed $T_{*}=2.332\times10^{6}\,\rm{K}$} 
\begin{tabular}{cccccccccccccccccccccc}
\hline
\hline
\centering
$B/\rm{G}=(B_{\rm{s}}/\rm{G},B_0/\rm{G})\,$ & $T_{*,\rm{in}}/\rm{K}\,$ & $L_{\rm{in}}/L_{\odot}$ & $L(T)/{\rm erg\ s^{-1}}$ & $T_{*,\rm{pr}}/\rm{K}$ \\ \hline
$(0, 0)$                & $2.332\times10^{6}$ & $1.00\times10^{-5}$ & $2.013\times10^{6}T^{3.500}$ & $2.223\times10^{6}$ \\ \hline
$(10^{11},5\times10^{14})$ & $2.332\times10^{6}$ & $1.26\times10^{-6}$ & $5.901*10^{-2}T^{4.541}$ & $2.317\times10^{6}$ \\ \hline
$(2\times10^{11},5\times10^{14})$ & $2.332\times10^{6}$ & $6.77\times10^{-7}$ & $2.996\times10^{-2}T^{4.545}$ & $2.324\times10^{6}$ \\ \hline
$(5\times10^{11},5\times10^{14})$ & $2.332\times10^{6}$ & $2.98\times10^{-7}$ & $1.317\times10^{-2}T^{4.545}$ & $2.328\times10^{6}$ \\ \hline
$(10^{12},10^{14})$ & $2.332\times10^{6}$ & $7.93\times10^{-7}$ & $3.715\times10^{-2}T^{4.541}$ & $2.323\times10^{6}$ \\ \hline
$(10^{12},5\times10^{14})$ & $2.332\times10^{6}$ & $1.60\times10^{-7}$ & $7.072\times10^{-3}T^{4.545}$ & $2.330\times10^{6}$ \\ \hline
$(2\times10^{12},10^{14})$ & $2.332\times10^{6}$ & $4.26\times10^{-7}$ & $1.882\times10^{-2}T^{4.545}$ & $2.327\times10^{6}$ \\ \hline
$(2\times10^{12},5\times10^{14})$ & $2.332\times10^{6}$ & $8.57\times10^{-8}$ & $3.474\times10^{-3}T^{4.552}$ & $2.331\times10^{6}$ \\ \hline
$(5\times10^{12},10^{14})$ & $2.332\times10^{6}$ & $1.87\times10^{-7}$ & $7.583\times10^{-3}T^{4.552}$ & $2.330\times10^{6}$ \\ \hline
$(5\times10^{12},5\times10^{14})$ & $2.332\times10^{6}$ & $3.76\times10^{-8}$ & $1.567*10^{-3}T^{4.550}$ & $2.332\times10^{6}$ \\ \hline
\hline
\label{table8}
\end{tabular}
\end{center}
\end{table*}

We find that an increase of the magnetic field strength results in a decrease in the coefficient $\alpha$ and increase in the exponent $\gamma$ in the $L=\alpha T^{\gamma}$ relation, as shown in Table \ref{table8}. 
Like the fixed $r_*$ case, the cooling rate decreases appreciably with an increase in magnetic field strength for $B_{\rm{s}} \ge 5\times10^{11}\,$G and $B_{0} \ge 5\times10^{14}\,$G.

\section{Discussion}
\label{sec5}
In this section, we discuss our results described in the previous sections and their physical significance.

\subsection{Non-magnetized white dwarfs}
\label{sec5.1}
From Table \ref{table1}, we see that as $L$ increases in the envelope, both $T_{*}$ and $\rho_{*}$ increase whereas $r_{*}$ decreases. This is owing to the fact that a white dwarf with a larger $T_{*}$ has more stored thermal energy, which it can radiate, giving rise to a larger $L$. Also, a larger $T_{*}$ corresponds to a larger $\rho_{*}$ by the EoS of non-degenerate matter, as seen from equation (\ref{eqn4}). For a fixed $T_{\rm{s}}$ and $R$, $r_{*}$ should decrease as $T_{*}$ increases. This is because the outer regions of the white dwarf are cooler than the inner ones. 

We also find that $|\Delta T/ \Delta r|$ = $|(T_{\rm s} - T_{*})/(R - r_{*})|$ and the cooling rate $|\Delta T/ \Delta t|$  = $|(T_{*,\rm{pr}} - T_{*,\rm{in}})/(t - t_{0})|$ increase with increase in luminosity of the white dwarf. 
Note that $L$ corresponds to the energy flux that is transported across a spherical surface and hence a larger luminosity means a larger flux (for a given radius) and a larger $\Delta T/ \Delta r$. From equation (\ref{eqn9}), 
it appears that hotter or more luminous white dwarfs cool faster because $T_{0}$ is larger. Therefore, the cooling rate should be faster for a white dwarf of larger luminosity.

\subsection{Magnetized white dwarfs of fixed interface radius}
\label{sec5.2}
In section \ref{sec3.1}, we have found how much the luminosity has to decrease for a magnetized white dwarf for it to have the same $r_*$ as a non-magnetized white dwarf. Then in section \ref{sec4.2}, we have also computed the cooling rates for the corresponding cases and used $L$ and $T_{\rm{s}}$ as obtained in section \ref{sec3.1} to estimate their evolution. Here we discuss our results.

In sections \ref{sec3.1} and \ref{sec4.2}, we have fixed $r_{*}$ and calculated $T_{*,\rm{in}}$, $T_{*,\rm{pr}}$, and $\rho_{*}$, and based
on this the present surface temperature could be determined.
We have used $r_{*} = 0.9978\,R$, which corresponds to $B = 0$ and $L = 10^{-5}\,L_{\odot}$. From Table \ref{table5}, we have seen that as $B$ increases, $\rho_{*}$ increases whereas $L$ and $T_{*}$ decrease for fixed $r_{*}$.

For the $B$ configuration that we have considered, the strength of the field increases with density. Therefore, $BdB/d\rho$ is positive and we obtain a smaller gradient $d\rho/dT$ for a given field strength for radially varying magnetic field as opposed to a radially constant (or zero) magnetic field (see equation \ref{eqn7}). Because the initial conditions are the same, we obtain a smaller $\rho$ at a given $T$ for a white dwarf with larger 
$B$, than $\rho$ at the same $T$ for a white dwarf with smaller $B$. Therefore, 
the presence of magnetic field suppresses the matter density at a given 
temperature compared to the non-magnetized case 
and thus we obtain a larger $T_{*}$ (see the right-hand panel of Fig. \ref{fig2}).

Now from equation (\ref{eqn8}), we have $dT/dr \propto \rho(\rho + \rho_{B})/B^2=\rho(\rho/B^2+1/8\pi c^2)$. However, a decrease in $\rho$ along with an increase in $B$ leads to a decrease in $dT/dr$ (see the right-hand panel of Fig. \ref{fig4}). Therefore, we have a smaller $T_{*}$ and a smaller $L$ for larger field strengths, for $r_{*}$ to be constant.

We find that $|\Delta T/\Delta r|$ and $|\Delta T/\Delta t|$ both decrease with $B$. As $T_{*}$ decreases with 
the increase in $B$ while $r_{*}$ remains fixed, a decrease in $|\Delta T/\Delta r|$ is expected. We know that $L$ is of the form $\alpha T^{\gamma}$ as given in Table \ref{table7}. Hence, we have
\begin{equation}
\tau \propto \frac{(T^{1-\gamma} - T_{0}^{1-\gamma})}{\alpha (\gamma - 1)}.
\label{eqn10}
\end{equation}
When $B$ increases, $\alpha (\gamma - 1)$ and $(T^{1-\gamma} - T_{0}^{1-\gamma})$ both decrease. However, the decrease in $\alpha (\gamma - 1)$ is more so that $\tau$ increases. With increasing $B$, $T_{0}$ and $\gamma$ do not change considerably whereas $\alpha$ decreases by orders of magnitude. Therefore, the cooling rate decreases with the increase in $B$.

\subsection{Magnetized white dwarfs of fixed interface temperature}
\label{sec5.3}
In section \ref{sec3.2}, we have computed the change of $L$ for a magnetized white dwarf of the same 
$T_*$ as a non-magnetized white dwarf. Then, in section \ref{sec4.3}, we have found the cooling rates for the corresponding cases and used $L$ and $T_{\rm{s}}$ as in section \ref{sec3.2} to obtain their evolution. Here we discuss our results.

In sections \ref{sec3.2} and \ref{sec4.3}, we have fixed $T_{*}$ and calculated $T_{*,\rm{in}}$, $r_{*}$, $\rho_{*}$ and $T_{*,\rm{pr}}$.
We have fixed $T_{*} = 2.332\times10^{6}\,$K, which is the interface temperature corresponding to $L = 10^{-5}\, L_{\odot}$ for the non-magnetic case and found that as $B$ increases, both $L$ and $r_{*}$ decrease, whereas $\rho_{*}$ increases, as can be seen from Table \ref{table6}. 

Because $\rho_{*} \propto T_{*}^{1/2}B_{\rm{s}}$ for a non-degenerate envelope, $\rho_{*}$ has to increase as $B_{\rm{s}}$ increases with $T_{*}$ fixed. Also, we know from section \ref{sec5.2} that the presence of magnetic field suppresses $\rho$ for a given $T$. The initial conditions for the $\rho - T$ profile are same, so we should have larger $d\rho/dT$ near the interface in the magnetic case. This happens because of a reduction in $L$ (equation \ref{eqn7}). Therefore, for $T_{*}$ to remain fixed with increasing field, $L$ must decrease. 

Now the initial conditions for the $T-r$ profile are the same as those for the non-magnetic case and $dT/dr$ near the surface is smaller for larger magnetic fields (from the right-hand panel of Fig. \ref{fig4}). So we obtain a smaller $r_{*}$ for a given $T_{*}$. 
We find that with increasing $B$, the luminosity is sufficiently small, in addition to $\rho$ being small. This counteracts the increase in $\rho_{B}$ making $dT/dr$ near the interface smaller. Therefore, $r_{*}$ decreases with increasing $B$ for fixed $T_{*}$. 

We find that the cooling rate $|\Delta T/\Delta t|$ decreases as magnetic field strength increases. The expression for the cooling time-scale is given by equation (\ref{eqn10}). In this case, the decrease in $\alpha(\gamma - 1)$ is larger than the decrease in $(T^{1-\gamma} - T_{0}^{1-\gamma})$. This makes $\tau$ larger for larger $B$.

\section{Summary and Conclusion}
\label{sec6}
We have investigated the effects of magnetic field on the luminosity and cooling of white dwarfs.
This is very useful to account for observability of recently proposed highly magnetized white dwarfs,
in particular those with central fields of $5\times 10^{14}\,$G. However, we have deferred our
investigation for white dwarfs with fields $B \gtrsim 10^{15}\,$G for future work. Such fields
affect the EoS significantly and might change the thermal conduction and observable properties more severely. 
It is important to note that magnetic fields in the
white dwarfs under consideration practically do not decay
by Ohmic dissipation and ambipolar diffusion during the
lifetime of the Universe \citep{heyl}. 
Even when the Hall drift plays
the dominant role in the decay of the magnetic field close to the white dwarf interface, the time-scale for an appreciable reduction is still about $1\,$Gyr for fields $10^{12} < B/\rm{G} < 10^{13}$ \citep{heyl}. Also various dynamo mechanisms cannot be ruled out to supplement
fields further.

We have computed the variation of luminosity of highly magnetized white dwarfs with magnetic field 
strength and evaluated the corresponding cooling time-scales for 
white dwarfs with the same fixed interface radius or temperature as their non-magnetic counterparts. 
We have found that at a given age of white dwarfs, the luminosity is suppressed with 
the increase in field strength, in 
addition to a marginal reduction of cooling rates. Therefore, white dwarfs with higher magnetic fields 
have lower luminosities and slower cooling, at the same interface radius or temperature, as
for non-magnetic white dwarfs. 

This apparent correlation between luminosity and magnetic field is found 
for higher fields only, $(B_{\rm{s}},B_0)\gtrsim (10^9,10^{13})\,$G.
At lower fields, there is neither any practical effect of magnetic fields
nor correlation. This is perfectly in accordance with observations so far,
as long as observed white dwarfs are assumed to have central field less
than $10^{13}\,$G. Indeed, there are very few white dwarfs observed so
far with $B_{\rm{s}} \approx 10^9\,$G. Interestingly, for $B_{\rm{s}} < 10^6\,$G,
observations suggest that higher field strength corresponds
to lower $T_{\rm{s}}$ and hence lower luminosity \citep{ferra}. 
From the number distribution of white dwarfs with field strength (\citealt{ferra}),
it can be seen that there are fewer white dwarfs observed with larger 
fields. Hence, extrapolating this
trend, we expect that our results would be in accordance with observations
when white dwarfs with higher field strength ($B_{\rm{s}} > 10^9\,$G) are observed.
As suggested by \citet{ferra}, non-detection of any apparent correlation 
between field and luminosity for $10^6 \lesssim  B/{\rm G} \lesssim 10^7$
may be due to the presence of possible effective bias while estimating parameters such as effective temperature
and gravity with models for non-magnetic white dwarfs. 
Although, there is a chance that biases could cancel each other out because we estimate 
temperatures using a wide range of methods, 
we simply cannot rule out that the effective biases are still there.

For a similar gravitational energy (similar mass and radius), an increasing magnetic energy
necessarily requires decreasing thermal energy for white dwarfs to be in equilibrium.
This results in a decrease in luminosity. Of course, understanding the evolution and structure of a white dwarf
is a complicated time-dependent nonlinear problem. Hence, our findings should be confirmed
based on more rigorous computations, without assuming beforehand the core
to be perfectly isothermal, self-similarity of the cooling process up to $10\,$Gyr, etc. Nevertheless, we have found that the luminosity could be as low as about $10^{-8}\,L_\odot$ for a white dwarf with the central field around $5\times 10^{14}\,$G and 
the surface field about $5\times10^{12}\,$G, for the same interface temperature as non-magnetic white dwarfs.
As a result, such white dwarfs appear 
to be invisible to
current astronomical techniques. However, with weaker surface fields, the luminosity tends to reach the observable limit. It is still about $10^{-6}\,L_\odot$ for surface fields of about
$B_{\rm{s}} \approx 10^9\,$G, with central fields $B_{0} \gtrsim 2\times 10^{13}\,$G. Note that 
the central field also plays an important role to determine luminosity. A lower $B_{0}$ makes the white dwarfs more observable for the same surface field.
Indeed, white dwarfs with surface fields $B_{\rm{s}} \approx 10^9\,$G are observed, whatever
be their number. We argue that such white dwarfs have relatively low central
fields. 
For a fixed interface radius, the luminosity could be much lower, $L \approx 10^{-9}\,L_\odot$, 
for central and surface fields of about $10^{12}\,$G and $5\times10^{11}\,$G, respectively. For surface fields approaching $10^9\,$G, $L \approx 10^{-8}\,L_\odot$, well below the observable limit, as long as central field $B_{0} \gtrsim 3\times10^{13}\,$G. 
Therefore, such white dwarfs, while expected to be present in the Universe, are virtually invisible to us, and perhaps lie in the lower left-hand corner in the Hertzsprung--Russell diagram.


\section*{Acknowledgments}
We thank Chanda J. Jog of IISc for discussion and continuous encouragement. 
We are highly indebted to the referee, Christopher Tout, for his careful
reading of the manuscript and for his numerous suggestions that substantially improved the paper.


\label{lastpage}

\end{document}